\documentclass[aps,prl,twocolumn,superscriptaddress,preprintnumbers,longbibliography,nofootinbib]{revtex4-1}

\usepackage{amssymb}
\usepackage{graphicx}
\usepackage{amsmath}
\usepackage{hyperref}
\usepackage{xspace}

\DeclareRobustCommand{\Eq}[1]{Eq.\,(\ref{#1})}

\usepackage{color}
\definecolor{darkblue}{rgb}{0,0,0.5}
\definecolor{darkred}{rgb}{0.5,0,0}
\definecolor{darkgreen}{rgb}{0,0.5,0}

\newcommand{\be}{\begin{equation}}
\newcommand{\ee}{\end{equation}}

\allowdisplaybreaks

\begin{document}

\preprint{}

\title{The Multiplicity Scaling of the Fragmentation Function}

\author{Zhong-Bo Kang}
\email{zkang@ucla.edu}
\affiliation{Department of Physics and Astronomy, University of California, Los Angeles, CA 90095, USA}
\affiliation{Mani L. Bhaumik Institute for Theoretical Physics, University of California, Los Angeles, CA 90095, USA}
\affiliation{Center for Frontiers in Nuclear Science, Stony Brook University, Stony Brook, NY 11794, USA}
\author{Robert Kao}
\email{rqk@ucla.edu}
\affiliation{Department of Physics and Astronomy, University of California, Los Angeles, CA 90095, USA}
\affiliation{Mani L. Bhaumik Institute for Theoretical Physics, University of California, Los Angeles, CA 90095, USA}
\author{Andrew J.~Larkoski}
\email{larkoski@ucla.edu}
\affiliation{Department of Physics and Astronomy, University of California, Los Angeles, CA 90095, USA}
\affiliation{Mani L. Bhaumik Institute for Theoretical Physics, University of California, Los Angeles, CA 90095, USA}

\begin{abstract} 
The single-particle inclusive fragmentation function and the particle multiplicity are observables of fundamental importance in studying properties of quantum chromodynamics at colliders.  It is well-known that at high energies, the multiplicity distribution satisfies KNO scaling in which all moments are proportional to powers of the mean multiplicity.  We prove that, under weak assumptions, the leading dependence of the fragmentation function on multiplicity is itself a kind of KNO scaling in which all moments are inversely proportional to powers of the mean multiplicity.  This scaling with multiplicity additionally accounts for the dominant dependence on collision energy in the fragmentation function.  The proof relies crucially on properties of the fragmentation function conditioned on the total multiplicity and application of the Stieltjes moment problem.  In the process, we construct a novel basis of the fragmentation function expressed as an overall exponential suppression times a series of Laguerre polynomials.  We study this scaling of the fragmentation function in experimental electron-position collision data and observe that residual scale violations are significantly reduced. 
\end{abstract}


\maketitle

A quantity of central importance in understanding the properties and behavior of the theory of the strong force, quantum chromodynamics (QCD), is the distribution of a particle's energy fraction that was produced in a collision event.  A whole chapter is devoted to this single-particle inclusive fragmentation function in the Particle Data Group's Review of Particle Physics \cite{ParticleDataGroup:2022pth} and its features in threshold limits, as a function of collision energy, and sensitivity to hadron species reveals a rich and varied structure and provides validation of the whole theoretical framework.  In the collisions of electrons and positrons, as we study in this Letter, particle $i$'s energy fraction $z_i$ is typically defined as \cite{Altarelli:1981ax,Ellis:1996mzs}
\begin{align}
z_i = \frac{2E_i}{\sqrt{s}}\,,
\end{align}
where $\sqrt{s}$ is the center-of-mass collision energy of the events.  This definition satisfies the energy conservation sum rule
\begin{align}
\sum_{i\in \text{event}} z_i = 2\,,
\end{align}
while momentum conservation restricts any individual energy fraction in the range $z_i \in [0,1]$.

With unit normalization, the fragmentation function $F(z)$ represents the probability that a particle in a particular event carries energy fraction $z$~\cite{Collins:1981uw}.  This is clearly related to the particle multiplicity $N$ of the event as a larger multiplicity would correspond to less energy per particle on average.  This can be made precise by introducing the fragmentation function conditioned on the multiplicity $N$, $F(z|N)$ \cite{Kang:2023ptt}.  The inclusive fragmentation function is defined as
\begin{align}
F(z) = \int dN\, F(z|N)\,p(N)\,,
\end{align}
where $p(N)$ is the probability distribution of multiplicity.  The mean of the conditional fragmentation function is fixed by the multiplicity,
\begin{align}\label{eq:condmean}
\langle z\rangle_N \equiv \int dz\, z\, F(z|N) = \frac{2}{N}\,.
\end{align}
We will always use the subscript $N$ to denote expectation values of the conditional fragmentation function.  

In this Letter, we establish general, robust properties of $F(z|N)$ that hold in the limit of large multiplicity, $N\to\infty$.  With these results, we will show that the leading dependence of the inclusive fragmentation function on multiplicity is reminiscent of the Koba-Nielsen-Olesen (KNO) scaling of particle multiplicity distribution \cite{Polyakov:1970lyy,Koba:1972ng}, where
\begin{align}
F(z) = \langle N\rangle\, \Phi\left(
\langle N\rangle z
\right)\,,
\end{align}
for a function $\Phi(x)$ that has no residual dependence on the mean multiplicity. Despite the extensive precise theoretical analysis, formulation, and predictions of the fragmentation function over its greater than 50 year history, including its perturbative scale evolution through the celebrated DGLAP equations \cite{Gribov:1972ri,Gribov:1972rt,Lipatov:1974qm,Dokshitzer:1977sg,Altarelli:1977zs}, we are unaware of any study that provided a quantitative, model-independent relationship between the multiplicity and the energy fraction of a single particle produced in a collision event.  Along the way to this goal, we construct a novel basis to represent the functional form of the fragmentation function that can provide a foundation for future theoretical studies, on which we can only briefly comment here.

In the limit of large multiplicity, the probability distribution of multiplicity $p(N)$ obeys KNO scaling, where
\begin{align}
p(N) = \frac{1}{\langle N\rangle}\,\psi\left(
\frac{N}{\langle N\rangle}
\right)\,,
\end{align}
where $\psi(x)$ is a universal function.  KNO scaling, and its violation, is well-established experimentally, e.g., at LEP \cite{DELPHI:1990ohs,ALEPH:1991ldi,DELPHI:1991qnt,OPAL:1991lzj,DELPHI:1996idl,ALEPH:1996oqp,DELPHI:1997nyf,DELPHI:2000ahn} and the LHC \cite{ALICE:2010cin,CMS:2010qvf,LHCb:2011jir,ALICE:2015olq,ALICE:2017pcy}.  This result of Eq.~\eqref{eq:condmean} then implies that the mean of the fragmentation function is
\begin{align}
\langle z\rangle = \int dz\, z\, F(z) =2 \langle N^{-1}\rangle\,.
\end{align}
With KNO scaling, the mean of the fragmentation function scales inversely proportional to the mean multiplicity, $\langle z\rangle \propto \langle N\rangle^{-1}$.  Without scaling properties of the conditional fragmentation function $F(z|N)$, we can not establish general relationships between higher moments of the fragmentation function and the multiplicity distribution. 

While one could establish properties of $F(z|N)$ through its perturbative scale evolution for example, here we will instead start from a set of weak assumptions and employ consistency relations and enforce qualities required of all probability distributions.  This procedure will result in a useful framework and functional form for calculations of the fragmentation function, which we leave to future work.  The assumptions that we employ here are:
\begin{enumerate}
\item The number of particles in the event $N$ is large, $N\to\infty$.

\item The distribution of particles is sufficiently smooth away from the boundaries of phase space.

\item Only particle energies are measured so all particles are treated as identical and indistinguishable.
\end{enumerate}

The conditional fragmentation function $F(z|N)$ can then be calculated from
\begin{align}
&F(z|N)\\
&\hspace{0.5cm}\propto \lim_{N\to\infty}\int_0 \prod_{i=1}^N [dz_i]\, f\left(z_1,\dotsc,z_N\right)\delta\left(
2-\sum_{i=1}^N z_i\nonumber
\right)\\
&\hspace{6cm}\times\delta(z-z_1)\,,\nonumber
\end{align}
where $f(z_1,z_2,\dotsc,z_N)$ is a positive semi-definite, integrable function that is permutation-symmetric in the particle energy fractions $z_i$.  Exploiting permutation symmetry, we select particle 1 to measure the energy fraction $z$.  Performing the integral over $z_1$ and introducing the variables $z_i \equiv x_i(2-z)$, this becomes
\begin{align}
&F(z|N)\propto e^{-\frac{N}{2}z} \lim_{N\to\infty} \int_0 \prod_{i=2}^N [dx_i]\\
&\hspace{1cm}\times f\left(z,x_2(2-z),\dotsc,x_N(2-z)\right)\delta\left(
1-\sum_{i=2}^N x_i
\right)\,,\nonumber
\end{align}
in the large $N$ limit.  The overall exponential factor restricts the region where there is significant support to $z\sim 2/N \ll 1$, so, to leading-order in $1/N$, we set $2-z \to 2$ in the argument of the function $f$:
\begin{align}
F(z|N)&\propto e^{-\frac{N}{2} z}\lim_{N\to\infty} \int_0 \prod_{i=2}^N [dx_i]\, f\left(z,2x_2,\dotsc,2x_N\right)\nonumber\\
&\hspace{3cm}\times\delta\left(
1-\sum_{i=2}^N x_i
\right)\,.
\end{align}
Note that the remaining $\delta$-function restricts the $x_i$ variables to lie in an $N-2$ simplex.

If the function $f\left(z,2x_2,\dotsc,2x_N\right)$ were analytic in its arguments about the origin, we could Taylor expand and ignore terms at linear order and beyond because they would be suppressed by powers of $z\sim 2/N$.  That is, in the large-$N$ limit, the conditional fragmentation function would reduce to
\begin{align}
F(z|N)\propto e^{-\frac{N}{2}z}\,,
\end{align}
the maximally-entropic distribution with a fixed mean $\langle z\rangle_N = 2/N$.  We should expect that the function $f\left(z,2x_2,\dotsc,2x_N\right)$ is non-analytic at the boundaries of phase space so this Taylor expansion about the origin will in general not be valid.  However, we can instead expand around the mean value of $\langle z\rangle_N = 2/N$, which, while small, is displaced from the boundary of phase space.  Generically, distributions should be smooth on the interior of phase space unless there were resonances or other new states, which here we do not consider.  

\begin{figure*}[t!]
\begin{center}
\includegraphics[width=0.45\textwidth]{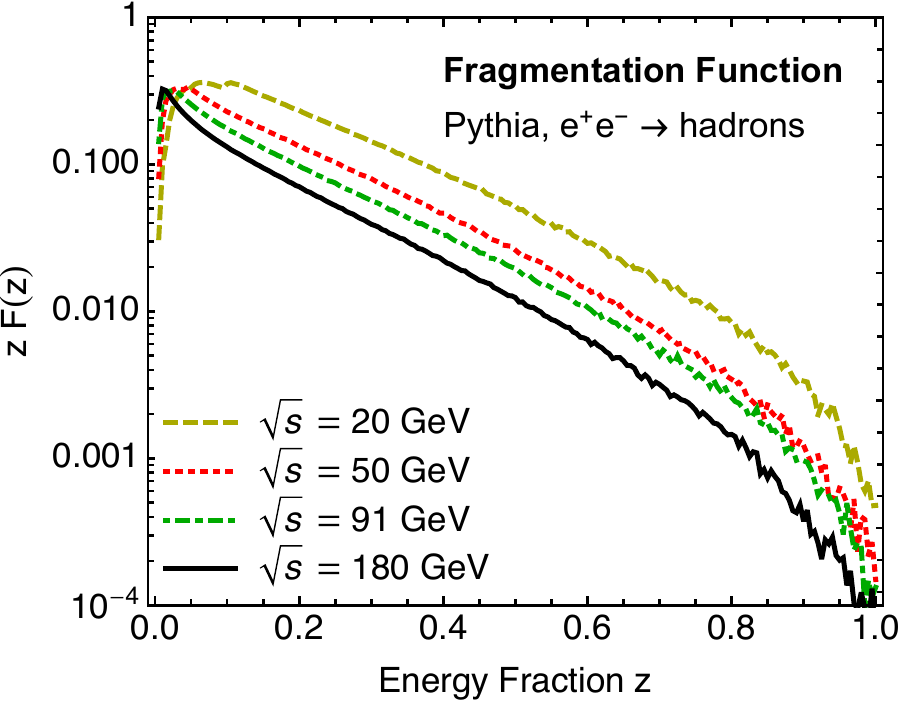}$\qquad$
\includegraphics[width=0.45\textwidth]{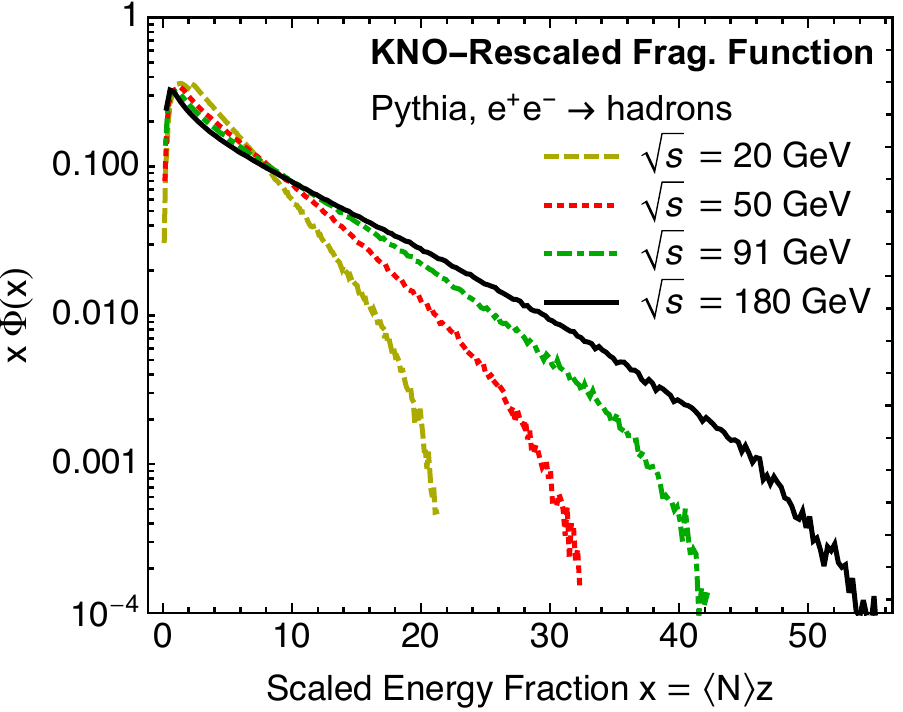}
\caption{\label{fig:frag}
The single-particle inclusive fragmentation function (left) in simulated $e^+e^-\to$ hadrons events and the corresponding KNO-rescaled fragmentation functions (right) at four different center-of-mass energies.
}
\end{center}
\end{figure*}

The exponential form of the conditional fragmentation function has effectively been considered in related contexts previously, e.g., Ref.~\cite{Wilk:2018kvg}.  However, this was studied within a particular model for multi-particle production, with explicit choices for the distributions of particle energies and multiplicity.  This and other studies therefore lacked a general framework or clear separation between fundamental hypotheses and constraints imposed by the form of model.

To perform the expansion away from the boundary, we introduce the variables
\begin{align}
x_i \equiv \frac{1}{N}+\epsilon_i\,,
\end{align}
and so the conditioned fragmentation function becomes
\begin{align}
&F(z|N)\\
&\propto e^{-\frac{N}{2} z}\lim_{N\to\infty}\!\!\! \int\limits_{-1/N} \prod_{i=2}^N [d\epsilon_i]\, f\left(z,\frac{2}{N}+2\epsilon_2,\dotsc,\frac{2}{N}+2\epsilon_N\right)\nonumber\\
&
\hspace{5cm}\times\delta\left(
\frac{1}{N}-\sum_{i=2}^N \epsilon_i
\right)\,.\nonumber
\end{align}
Now, by permutation symmetry and the remaining $\delta$-function, the expectation value of any of the $\epsilon_i$ variables is $\langle \epsilon_i\rangle_N \to 1/N^2$, in the large $N$ limit.  Thus again, to leading order in $1/N$, we can set $\frac{1}{N}+\epsilon_i \to \frac{1}{N}$.  Therefore, to leading-order in $1/N$, the conditioned fragmentation function can be expressed as
\begin{align}
F(z|N)&\propto e^{-\frac{N}{2} z}\, f\left(z,\frac{2}{N},\dotsc,\frac{2}{N}\right)\\
&\equiv e^{-\frac{N}{2} z}\, f\left(\frac{N}{2}z\right)\,.\nonumber
\end{align}
At right, we have concatenated the function into a function of a single argument, $f(Nz/2)$.  The argument must be of the form $Nz\sim 1$, otherwise we could expand the function about $z\sim 2/N$ and ignore terms that depend on the displacement from $2/N$ to leading order in $1/N$.  However, there may be additional dependence on the multiplicity $N$ in the function $f(Nz/2)$, which we address as follows.

The conditioned fragmentation function can be expanded in terms of Laguerre polynomials as
\begin{align}\label{eq:lagexpx}
F(x|N) = e^{-x}\, \left[
1+\sum_{n = 2}^\infty c_n\, L_n(x)
\right]\,,
\end{align}
where $x = Nz/2$, is unit normalized, and has unit mean, $\langle x\rangle_N = 1$.  The Laguerre polynomials are familiar as the radial energy eigenstate wavefunctions of the hydrogen atom and are defined to be
\begin{align}
L_n(x) = \sum_{k=0}^n {n\choose k} \frac{(-1)^k}{k!}\, x^k\,,
\end{align}
and are orthonormal with respect to the exponential kernel
\begin{align}
\delta_{mn} = \int_0^\infty dx\, e^{-x}\, L_m(x)\, L_n(x)\,.
\end{align}

In general, it might seem that the coefficients $c_n$ in \Eq{eq:lagexpx} can have dependence on multiplicity $N$, but we will show that any dependence in the $N\to\infty$ limit leads to a pathological expansion.  By the orthogonality of the Laguerre polynomials, the coefficients $c_n$ can be isolated by taking moments of $F(x|N)$.  These moments can be evaluated to find
\begin{align}\label{eq:laguerremom}
\langle x^l\rangle_N  = l!\left(
1+\sum_{n=2}^l (-1)^n\,{l\choose n}\, c_n
\right)\,.
\end{align}
To ensure that $F(x|N)$ is a probability distribution, namely, positive and integrable on $x\in[0,\infty)$, there are strong constraints imposed on the moments, as established by the Stieltjes moment problem \cite{stieltjes1894recherches}.  This was also recently exploited to establish positivity constraints on general effective Lagrangians \cite{Arkani-Hamed:2020blm}.

The solution to the Stieltjes moment problem is that the Hankel matrix of moments must be completely positive.  For brevity here, we will only explicitly consider the lowest dimension non-trivial Hankel matrices and just comment on generalization to higher dimensions that include more moments.  To ensure positivity, the $2\times 2$ Hankel matrices
\begin{align}
\Delta_1 \equiv \left(
\begin{array}{cc}
\langle x^0\rangle_N & \langle x^1\rangle_N\\
\langle x^1\rangle_N & \langle x^2\rangle_N\\
\end{array}
\right)\,,
\end{align}
and
\begin{align}
\Delta_1^{(1)} \equiv \left(
\begin{array}{cc}
\langle x^1\rangle_N & \langle x^2\rangle_N\\
\langle x^2\rangle_N & \langle x^3\rangle_N\\
\end{array}
\right)\,,
\end{align}
must have positive determinants, $\det \Delta_1 > 0$ and $\det \Delta_1^{(1)} > 0$.  

\begin{figure*}[t!]
\begin{center}
\includegraphics[width=0.45\textwidth]{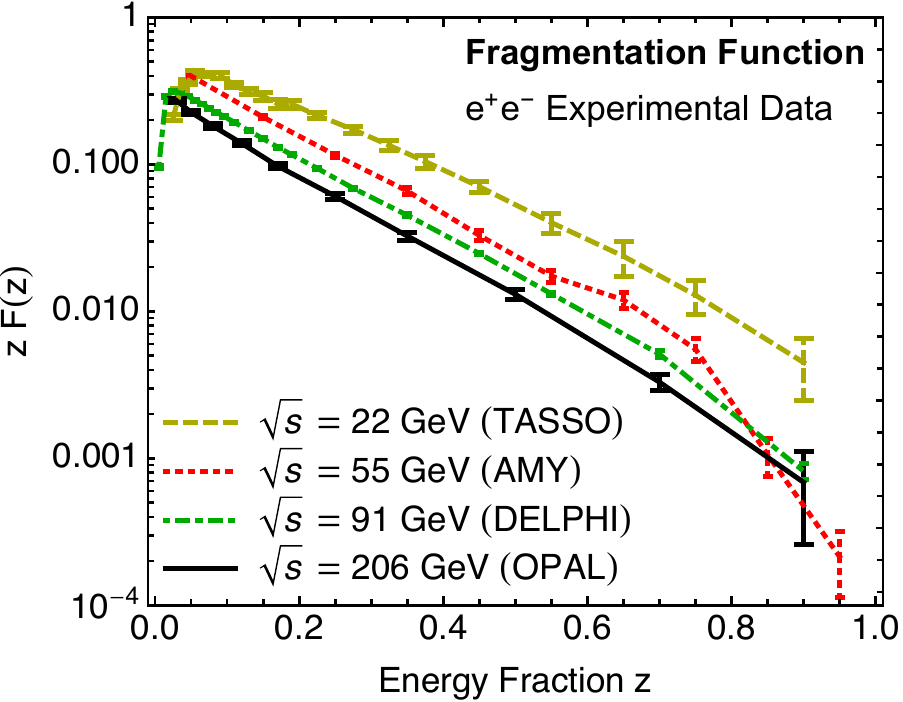}$\qquad$
\includegraphics[width=0.45\textwidth]{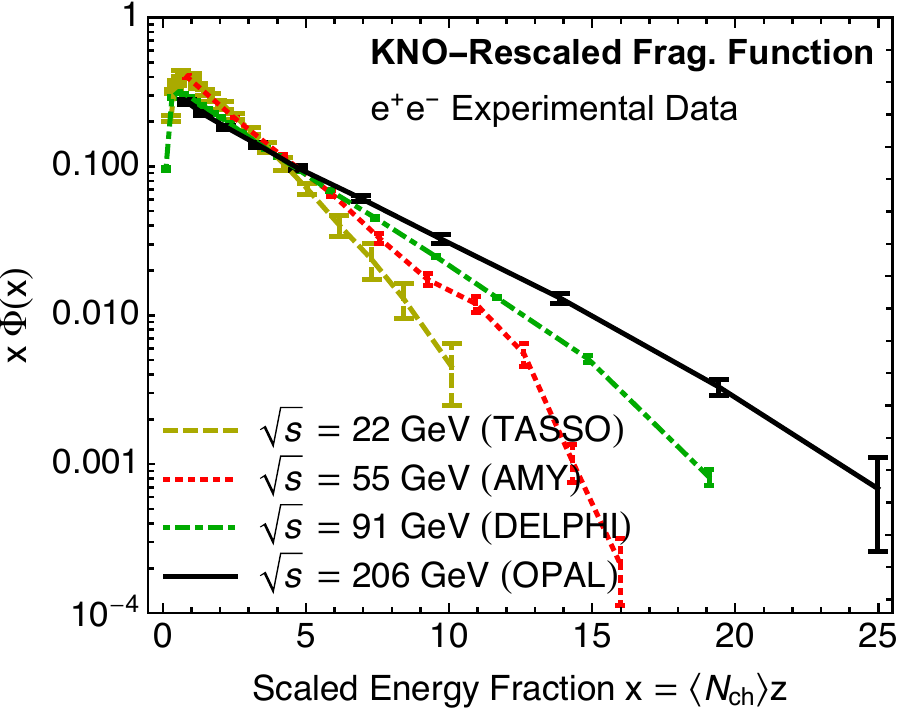}
\caption{\label{fig:fragdata}
The single-particle inclusive fragmentation function (left) in $e^+e^-\to$ hadrons experimental data and the corresponding KNO-rescaled fragmentation functions (right) at four different center-of-mass energies.
}
\end{center}
\end{figure*}

The determinants of the Hankel matrices and the Stieltjes constraints, using \Eq{eq:laguerremom}, are then
\begin{align}
\det \Delta_1 =  2c_2+1> 0\,,
\end{align}
and 
\begin{align}
\det\Delta_1^{(1)} = 6+18 c_2-6c_3-4(1+c_2)^2> 0\,.
\end{align}
If we assume that the coefficient $c_2\sim N^a$, for $a>0$ as $N\to\infty$, then the constraint on $\Delta_1$ can easily be satisfied.  However, with this assumption, the constraint on $\Delta_1^{(1)}$ then requires, in the large $N$ limit,
\begin{align}
\det\Delta_1^{(1)}\to -6c_3-4c_2^2> 0\,,
\end{align}
or that
\begin{align}
c_3 \lesssim -N^{2a}\,.
\end{align}
Therefore, to ensure that $F(x|N)$ is a probability distribution, the expansion coefficient $c_3$ must scale with multiplicity $N$ parametrically larger than $c_2$.  One finds that $c_4$ must correspondingly scale parametrically larger with $N$ than $c_3$, and this pattern continues to higher coefficients.  Such an expansion is terribly non-convergent, which violates our assumption that distributions on the interior of phase space are sufficiently smooth.  Therefore, there can be no explicit dependence on the multiplicity $N$ in the conditioned fragmentation function of \Eq{eq:lagexpx}; in the large $N$ multiplicity limit, the coefficients $c_n$ are independent of $N$.  These coefficients may still depend on other, suppressed, parameters, like the collision energy or particle masses, but we leave that study for future work.

This result implies that this conditional fragmentation function satisfies a KNO-like scaling, where
\begin{align}
F(z|N) = \frac{1}{\langle z\rangle_N}\, \mathcal{F}\left(\frac{z}{\langle z\rangle_N}\right) = \frac{N}{2}\,\mathcal{F}\left( \frac{N}{2}z\right)\,,
\end{align}
and the function $\mathcal{F}(x)$ is independent of multiplicity $N$ in the limit $N\to\infty$.  This result enables direct calculation of the leading dependence of the fragmentation function on the multiplicity distribution.  In this limit, we have established the scaling properties of the conditional fragmentation function $F(z|N)$, and the multiplicity distribution $p(N)$ satisfies KNO scaling.  We have
\begin{align}\label{eq:fragrescale}
F(z) &\to \int dN\, \frac{N}{2} \mathcal{F}\left(\frac{N}{2}z\right)\, \frac{1}{\langle N\rangle}\,\psi\left(
\frac{N}{\langle N\rangle}
\right)\\
&=\frac{\langle N\rangle}{2} \int dy \,y\,\mathcal{F}\left(\langle N\rangle z \frac{y}{2}\right) \, \psi(y)\nonumber\\
&\equiv \langle N\rangle\, \Phi\left(\langle N\rangle z\right)\,,\nonumber
\end{align}
for a function $\Phi(x)$ that has no residual dependence on mean multiplicity $\langle N\rangle$.  $\Phi(x)$ may still depend on the collision energy, but only in a way that is independent of $\langle N\rangle$.

We can observe this scaling of the fragmentation function in simulated data, which is displayed in Fig.~\ref{fig:frag}.  We generated $e^+e^-\to$ hadrons collision events in Pythia 8.309 \cite{Bierlich:2022pfr} at center-of-mass collision energies of $\sqrt{s} = 20,50,91,180$ GeV.  Here, we plot the single particle inclusive fragmentation function multiplied by the energy fraction $z F(z)$ at left.  As anticipated, at higher energies and therefore higher average multiplicities, the fragmentation function becomes more and more strongly peaked near $z = 0$.  However, by instead plotting the rescaled fragmentation function of Eq.~\eqref{eq:fragrescale}, at right in Fig.~\ref{fig:frag}, the broad spread between the distributions at different energies is dramatically reduced, especially at small values of the parameter $x=\langle N\rangle z$, denoted as ``scaled energy fraction'' in the figure.  Sample dependence is observed at the upper limits, where the rescaled fragmentation function is most sensitive to the deviations from the asymptotic $N\to\infty$ limit.  Further validation studies of the scaling behavior of the conditioned fragmentation function are presented in the Supplemental Material. 

With data from $e^+e^-$ collider experiments, we can test this leading scaling behavior of the fragmentation function directly.  To do this comparison, we use four data sets at collision energies comparable to the energies used in simulated data.  The data sets we use here are, with their measured average charged particle multiplicities:
\begin{itemize}
\item TASSO at $\sqrt{s}=22$ GeV \cite{TASSO:1990cdg,TASSO:1983cre}, $\langle N_\text{ch}\rangle = 11.22\pm 0.07$,

\item AMY at about $\sqrt{s}=55$ GeV \cite{AMY:1989feg,AMY:1990hyi}, $\langle N_\text{ch}\rangle = 16.82\pm 0.22$,

\item DELPHI at $\sqrt{s}=91$ GeV  \cite{DELPHI:1997oih}, $\langle N_\text{ch}\rangle = 21.21\pm 0.2$,

\item OPAL at $\sqrt{s}=206$ GeV \cite{OPAL:2004prv}, $\langle N_\text{ch}\rangle = 27.75\pm 0.67$.
\end{itemize}
Assuming that isospin (or flavor symmetry) is approximately conserved, the mean charged particle multiplicity just differs by a constant factor from the mean total multiplicity, independent of energy scale.  The fragmentation functions measured in these data are plotted in Fig.~\ref{fig:fragdata}.  At left, are the raw fragmentation functions that exhibit a structure that peaks more sharply near $z=0$ as collision energy or multiplicity increases.  At right, we rescale the fragmentation function by the mean charged particle multiplicity and observe that the scale dependence is reduced as multiplicity or collision energy increases, experimentally demonstrating this scaling relation of the fragmentation function.

We have established the leading scaling law for the single-particle inclusive fragmentation function with the mean particle multiplicity.  The proof of this scaling required construction of a systematic expansion of the fragmentation function conditioned on multiplicity in terms of Laguerre polynomials and constraints imposed on its coefficients by the Stieltjes moment problem.  Residual dependence on collision energy is then largely uncorrelated with the mean multiplicity and enables isolation of other physical effects that control the features of the fragmentation function.  
This general analysis and dependence on multiplicity may provide insight into features of the fragmentation function, from the Gaussian peak's dependence on energy in the distribution of $\log1/z$ \cite{Dokshitzer:1982xr,ParticleDataGroup:2022pth,Ellis:1996mzs} to the observed transition between recombination and fragmentation of partons in high-multiplicity heavy ion collisions \cite{Fries:2003vb,Greco:2003xt}.
We look forward to explorations and refinements of this analysis in the future.

\acknowledgments

A.L.~thanks Tao Han, Alexander Karlberg, Leif L\"onnblad, Michelangelo Mangano, Pier Monni, Marek Sch\"onherr, Andrzej Siodmok, Dave Soper, Alba Soto-Ontoso, Gregory Soyez, and Gherardo Vita for discussions.  We thank Barbara Jacak and Farid Salazar for suggestions on presentation of the scaling properties.  This work is supported by the National Science Foundation under grant No.~PHY-1945471.  This work was supported in part by the UC Southern California Hub, with funding from the UC National Laboratories division of the University of California Office of the President.

\bibliography{mult}

\end{document}


\title{The Multiplicity Scaling of the Fragmentation Function: Supplemental Material}

\author{Zhong-Bo Kang}
\email{zkang@ucla.edu}
\affiliation{Department of Physics and Astronomy, University of California, Los Angeles, CA 90095, USA}
\affiliation{Mani L. Bhaumik Institute for Theoretical Physics, University of California, Los Angeles, CA 90095, USA}
\affiliation{Center for Frontiers in Nuclear Science, Stony Brook University, Stony Brook, NY 11794, USA}
\author{Robert Kao}
\email{rqk@ucla.edu}
\affiliation{Department of Physics and Astronomy, University of California, Los Angeles, CA 90095, USA}
\author{Andrew J.~Larkoski}
\email{larkoski@ucla.edu}
\affiliation{Department of Physics and Astronomy, University of California, Los Angeles, CA 90095, USA}

\maketitle

\begin{figure*}[b]
\begin{center}
\includegraphics[width=0.465\textwidth]{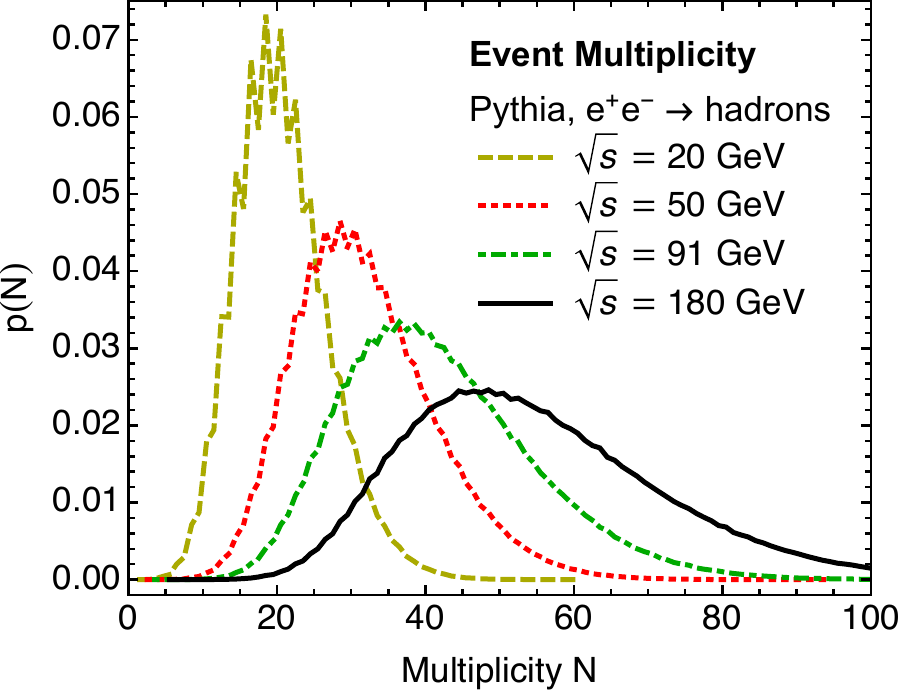}$\qquad$
\raisebox{-0.5cm}{\includegraphics[width=0.45\textwidth]{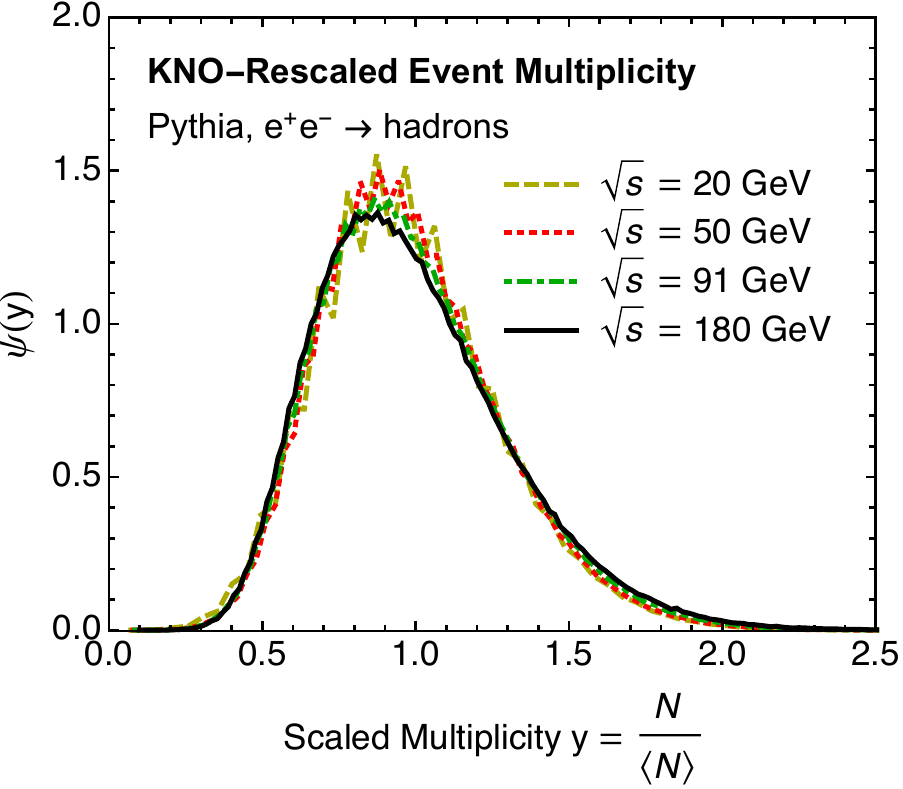}}
\caption{\label{fig:mult}
The particle multiplicity distributions (left) in simulated $e^+e^-\to$ hadrons events and the corresponding KNO-rescaled distributions (right) at four different center-of-mass energies.
}
\end{center}
\end{figure*}

Here, we present Supplemental Material for the article that further validates the predictions made from our scaling analysis.  In Fig.~\ref{fig:mult}, we demonstrate KNO scaling with the same simulated data as used in the main article.  As the energy increases, the particle multiplicity correspondingly increases, but when the distributions are rescaled by their respective means, they collapse to a nearly-identical, universal, distribution.

We focus on validation of the main result of the article, that the conditioned fragmentation function $F(z|N)$ is purely a function of $zN$ in the large multiplicity $N$ limit (see Eq.~(25) of the article):
\begin{align}
\lim_{N\to\infty}F(z|N) = \frac{N}{2}{\cal F}\left(\frac{N}{2}z\right)\,,
\end{align}
where ${\cal F}(x)$ with $x = Nz/2$ is independent of $N$.  This relationship implies that all moments in $z$ of $F(z|N)$ are inversely proportional to powers of multiplicity $N$:
\begin{align}
\langle z^n\rangle_N = \int dz\, z^n\, F(z|N)\propto N^{-n}\,,
\end{align}
for any $n$.  To test this, we will study the scaled moments $\langle x^n\rangle_N$, which should be independent of $N$ at sufficiently high multiplicity.

\begin{figure}[t]
\begin{center}
\includegraphics[width=0.45\textwidth]{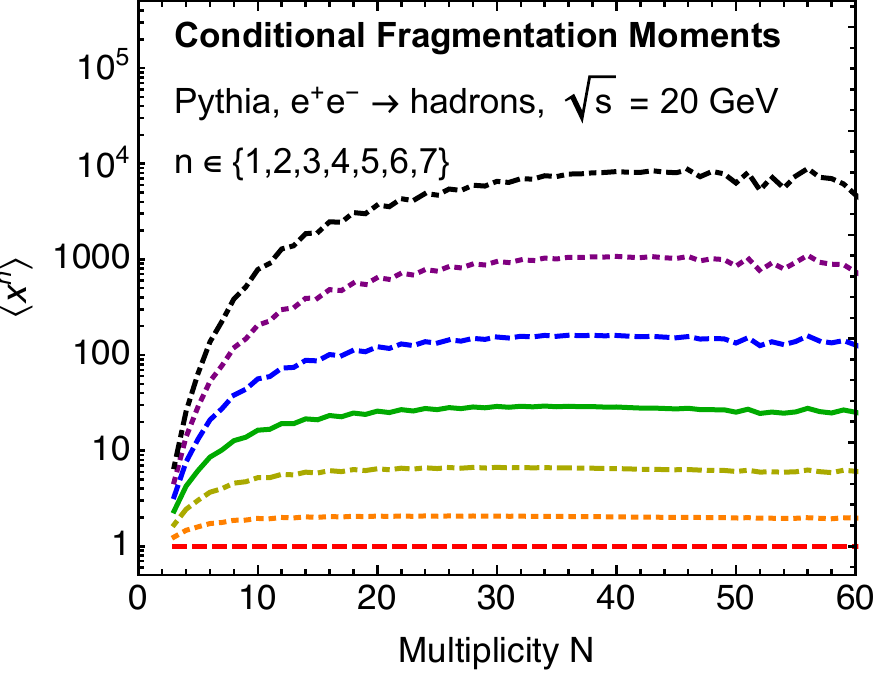} $\qquad$
\includegraphics[width=0.45\textwidth]{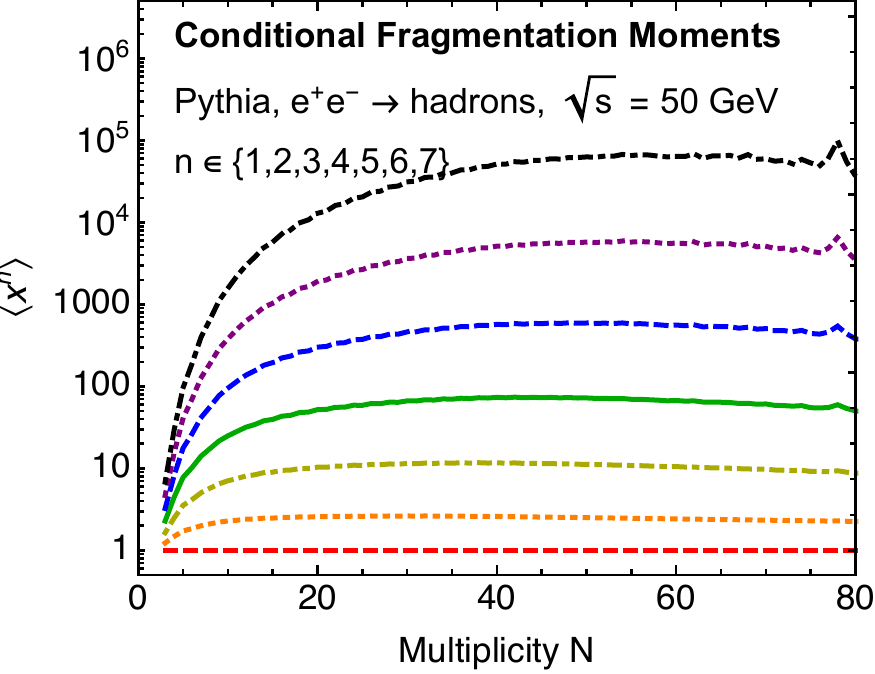} \\
\vspace{0.5cm}
\includegraphics[width=0.45\textwidth]{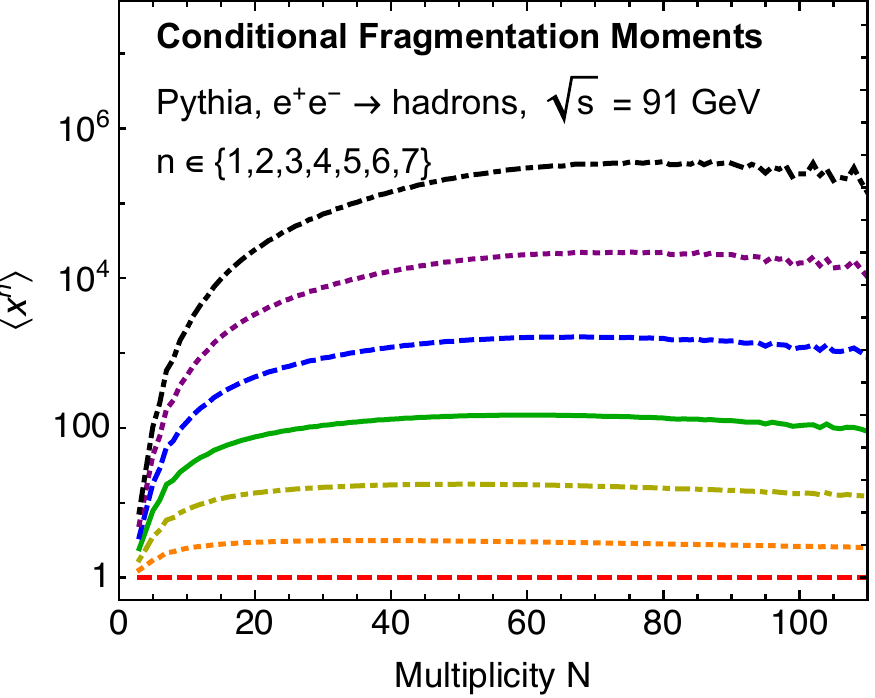} $\qquad$
\includegraphics[width=0.45\textwidth]{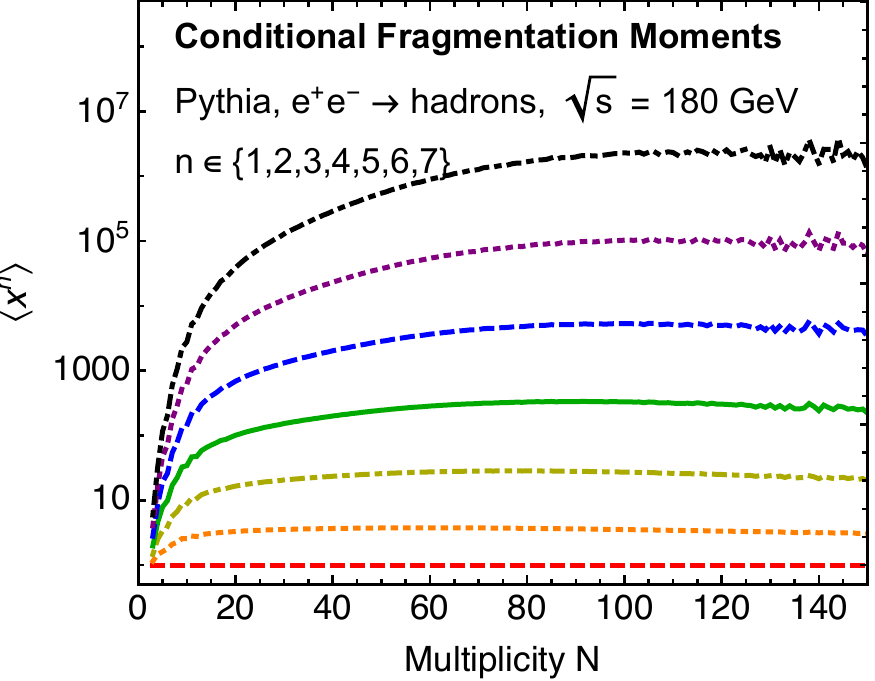}
\caption{\label{fig:pymoms}
Plots of the scaled moments $\langle x^n\rangle_N$, where $x= Nz/2$, for $n=1,2,3,4,5,6,7$ in increasing value measured in $e^+e^-\to $ hadrons events generated in Pythia at center-of-mass energies of $\sqrt{s}=20,50,91,180$ GeV.
}
\end{center}
\end{figure}

In the simulated $e^+e^-\to$ hadrons event samples from Pythia, we generate 50 million events at each center-of-mass collision energy $\sqrt{s} = 20,50,91,180$ GeV.  On these events, we measure the moments $\langle x^n\rangle_N$ for $n=1,2,3,4,5,6,7$ and bin them in multiplicity $N$.  The result of this analysis is presented in Fig.~\ref{fig:pymoms}.  At all energies, there is a steep rise in the moments at small multiplicity, but then the moments plateau, demonstrating to very good approximation that they do indeed lose dependence on multiplicity $N$ at sufficiently large multiplicity.  Some residual $N$ dependence is observed; at the very highest multiplicities, the value of the moments decrease slightly, but this is a small effect over a wide range of multiplicities.  Therefore, these simulated data do indeed respect the scaling of the conditioned fragmentation function $F(z|N)$ established in the main article.

